\documentclass[prd, showpacs, showkeys, twocolumn, floatfix]{revtex4}

\usepackage{graphicx}
\usepackage{amsmath, amsfonts, amssymb, bm}

\usepackage{slashed}
\usepackage{dsfont}

\usepackage[]{psfrag}

\psfragscanoff
\begin{document}

\title{Spin correlations in nonperturbative electron-positron pair creation by petawatt laser pulses colliding with a TeV proton beam}

\author{Tim-Oliver M\"uller$^{1,2}$}
\author{Carsten M\"uller$^1$}
\affiliation{$^1$Max-Planck-Institut f\"ur Kernphysik, Saupfercheckweg 1, D-69117 Heidelberg, Germany\\
$^2$Physik-Department, Technische Universit\"{a}t M\"{u}nchen, D-85747
Garching, Germany}

\date{\today}

\begin{abstract}
The influence of the electron spin degree of freedom on nonperturbative electron-positron pair production by high-energy proton impact on an intense laser field of circular polarization is analyzed. Predictions from the Dirac and Klein-Gordon theories are compared and a spin-resolved calculation is performed. We show that the various spin configurations possess very different production probabilities and discuss the transfer of helicity in this highly nonlinear process. Our predictions could be tested by combining the few-TeV proton beam at CERN-LHC with an intense laser pulse from a table-top petawatt laser source.
\end{abstract}


\maketitle

\section{Introduction}

The interaction of photons with spin-polarized electrons has been studied for a long time, with a focus on high-energy bremsstrahlung and Bethe-Heitler pair production in the field of an atomic nucleus. These investigations provided deep insights into the subtle quantum nature of helicity transfer in scattering reactions \cite{McVoy,Olsen}. Today, Bethe-Heitler pair production by circularly polarized, high-energy photons is utilized for generation of positron beams with high degree of polarization, which are of interest for future experiments in particle physics \cite{beam}.

When the photon source is an intense laser field, nonlinear generalizations of photo-induced processes may take place, involving the absorption (or emission) of more than one laser photon. Electron scattering processes in optical laser fields of ultra-high intensity $\gtrsim 10^{18}$\,W/cm$^2$ have been studied extensively in recent years, both theoretically \cite{Salamin,Ehlotzky} and experimentally \cite{exp}. Such fields are characterized by a large value of the parameter $\xi\equiv eE_0/(m\omega_0)$ of order unity, giving rise to pronounced relativistic effects. Here, $E_0$ and $\omega_0$ are the laser peak field strength and frequency, and $-e$ and $m$ are the electron charge and mass, respectively. Note that natural units with $\hbar = c = \epsilon_0 = 1$ are used.

The influence of the electron spin in relativistic multiphoton scattering processes in strong laser fields has been investigated with respect to atomic photoionization \cite{Faisal}, Compton and Mott \cite{Ehlotzky,Mott} scattering, and $e^+e^-$ pair production in photon-multiphoton collisions (strong-field generalization of the Breit-Wheeler process) \cite{Tsai,Serbo}. The spin effects were generally found to be rather small.

Pronounced spin signatures, however, may be expected at even larger field intensities corresponding to a sizeable value of the parameter $\chi\equiv\xi\omega_0/m\sim 1$ \cite{Uggerhoj,Walser}. It is interesting to note that the latter condition is equivalent to $E_0\sim E_S$, with the Schwinger field $E_S=m^2/e$ which sets the natural scale for field-induced pair production from vacuum \cite{Schwinger}. Indeed, characteristic differences between fermionic and bosonic particles have been revealed theoretically with respect to pair creation in an oscillating electric field \cite{Popov} and in recent studies of the Klein paradox \cite{Grobe}.

The highest laser field strengths available in the laboratory today reach $E_0\sim 10^{-4}E_S$ \cite{Mourou}. They can be amplified effectively in collisions of laser pulses with relativistic particle beams. When, for example, a few-TeV proton beam is brought into collision with a petawatt laser pulse of $\gtrsim 10^{21}$\,W/cm$^2$ intensity, the effective peak laser field strength in the proton rest frame closely approaches the Schwinger value due to a large Doppler boost (see \cite{SLAC} for a similar experimental setup utilizing the SLAC electron beam). 

The recent commissioning of the LHC at CERN \cite{LHC} has therefore stimulated a sustained interest from theoreticians in strong-field QED processes, mainly $e^+e^-$ pair production, occuring in ultrarelativistic proton-laser collisions \cite{MVG,Kaminski,Kuchiev,Milstein,PLB,ADP1,ADP2}. Besides, table-top petawatt laser systems are currently being developed \cite{Krausz} which could, in principle, be operated in conjunction with the LHC. The pairs would be generated by a multiphoton generalization of the Bethe-Heitler process: 
\begin{eqnarray}
n\omega + \gamma^* \to e^+e^-\,,
\end{eqnarray}
where $n$ is the photon order, which may vary over a broad range, $\omega$ is the Doppler-shifted laser frequency, and $\gamma^*$ denotes a virtual photon from the Coulomb field of the proton (some low-order Feynman diagrams of this process are depicted in Fig.\,1 of \cite{PLB}). In the strong-field Bethe-Heitler reaction, huge amounts of energy and momentum are imparted into the QED vacuum, corresponding to the absorption of $n\sim 10^5$--$10^6$ laser photons. Detailed predictions of total production rates as well as energy and angular positron spectra have been provided \cite{MVG,Kaminski,Kuchiev,Milstein,PLB,ADP1}. Spin-resolved results, however, were not reported yet.

Interesting questions are connected with the role of the electron spin in the nonlinear Bethe-Heitler process (1): How does the spin degree of freedom affect the total production rate? Which electron-positron spin configurations are favored? How does the helicity transfer by thousands of absorbed laser photons compare with the helicity transfer by a single photon of the same (total) energy?

In the present Letter, we provide answers to these questions by analyzing the influence of the electron spin on strong-field Bethe-Heitler pair production. The parameter regime with $\xi\gg 1$ and $\chi\sim 1$ is considered, where the coupling to the external laser field is manifestly nonperturbative. First, we compare the predictions of the Dirac and Klein-Gordon theories for spin-$\frac{1}{2}$ and spinless particles, respectively. Moreover, we provide explicitely spin-resolved production rates by applying the Dirac spin projector formalism within the framework of laser-dressed QED. In particular, we show that the dominant contribution to the total rate stems from antiparallel spin orientations of the electron and positron, whereas the smaller contribution from parallel lepton spins can be associated with the Klein-Gordon rate prediction.

For further recent studies on photo- and laser-induced $e^+e^-$ pair creation we refer the reader to \cite{Kuraev,Novak,Ruf,Gies,Tuchin,Bulanov,Marklund}.

\section{Comparison of the spinor and scalar cases}

In order to quantify the influence of the electron spin degree of freedom on the total rate of strong-field Bethe-Heitler pair creation, we first compare the corresponding predictions from the Dirac and Klein-Gordon theories. Similar comparative studies have been carried out with respect to Compton and Mott scattering \cite{Ehlotzky}.

Within the usual theoretical approach to pair creation in combined laser and nuclear Coulomb fields, which correctly accounts for the spinor character of the leptons, the process amplitude in the proton rest frame reads \cite{MVG,Kaminski}
\begin{eqnarray}
\label{S}
{\mathcal M}_{p_+p_-}^{s_+s_-} = -ie\int d^4x\,[\Psi_{p_-,s_-}^{(-)}]^\dagger V(r)\Psi_{p_+,s_+}^{(+)}.
\end{eqnarray}
The leptons are described by Dirac-Volkov states $\Psi_{p_\pm,s_\pm}^{(\pm)}$ which include their interaction with the plane-wave laser field to all orders (Furry picture). The states are labeled by the free four-momenta $p_\pm^\mu=({p_\pm^0,\boldsymbol p}_\pm)$ and spin quantum numbers $s_\pm$ outside the field. The indices $+$ and $-$ refer to the sign of the particle's electric charge. The Coulomb potential of the proton $V(r)=e/(4\pi r)$ is taken into account within the first-order of perturbation theory. Note that the nuclear field at a distance of a Compton wave length amounts to $\sim 10^{14}$\,V/cm which is substantially smaller than the near-critical laser field. 

The Dirac-Volkov states represent exact solutions of the Dirac equation in the presence of a classical plane electromagnetic wave. They are given by \cite{LL}
\begin{eqnarray}
\label{Volkov}
\Psi^{(\pm)}_{p_\pm,s_\pm} = \sqrt{\frac{m}{p^0_\pm}} \left(1\mp\frac{e\slashed{k}\slashed{A}}{2(kp_\pm)}\right) u^{(\pm)}_{p_\pm,s_\pm}\,{\rm e}^{i\Lambda^{(\pm)}}
\end{eqnarray}
with
\begin{eqnarray}
\label{Volkovphase}
\Lambda^{(\pm)} = \mp (p_\pm x) - {1\over (kp_\pm)}\int^\eta 
\left[ e{\boldsymbol p_\pm\cdot \boldsymbol A}(\tilde\eta) + {e^2\over 2}{\boldsymbol A}^2(\tilde\eta) \right]d\tilde\eta
\end{eqnarray}
corresponding to the classical action of the leptons in the laser field. Here, the Minkowski product of two four-vectors $a^\mu=(a^0,\mbox{\boldmath$a$})$ and $b^\mu=(b^0,\mbox{\boldmath$b$})$ is denoted as $(ab) = a^0b^0-\mbox{\boldmath$a\cdot b$}$. 
Especially $\slashed{a}=(\gamma a)=\gamma^\mu a_\mu$, with the Dirac matrices $\gamma^\mu$.
The four-potential of the laser field $A^\mu=(0,{\boldsymbol A})$ is taken in the radiation gauge. For definiteness, we will always assume a circularly polarized field of the form ${\boldsymbol A}(\eta)=A_0({\bf e}_x \cos\eta+{\bf e}_y \sin\eta)$, with the laser phase $\eta=(kx)=\omega t-{\boldsymbol{k\cdot r}}$. Note that this field polarization corresponds to right-handed laser photons. 

The Dirac-Volkov states obey the usual normalization condition of continuum states to a three-dimensional $\delta$ function in the particle three-momentum. The total rate for pair production is obtained from the square of the amplitude \eqref{S}, integrated over the outgoing particle momenta and summed over their spins,
\begin{eqnarray}
\label{R}
R = \sum_{s_+,s_-} R^{s_+s_-},\  R^{s_+s_-}=\int \frac{d^3p_+}{(2\pi)^3} \int \frac{d^3p_-}{(2\pi)^3} |{\mathcal M}_{p_+p_-}^{s_+s_-}|^2.
\end{eqnarray}
The spin-summed square of the amplitude can be transformed into a trace over Dirac matrices in the usual way.

If the electron was a spinless particle, its time evolution would be governed instead by the Klein-Gordon equation. The corresponding scalar Volkov solution in the external laser field would read \cite{Ehlotzky}
\begin{eqnarray}
\label{VolkovKG}
\Phi_{p_\pm} = \frac{1}{\sqrt{2p_\pm^0}}\,{\rm e}^{i\Lambda^{(\pm)}}
\end{eqnarray}
where $\Lambda^{(\pm)}$ is given again by Eq.\,(\ref{Volkovphase}). They obey the same normalization condition as the Dirac-Volkov states of Eq.\,\eqref{Volkov}. In the scalar case, the amplitude describing the creation of a pair of spinless particles with four-momenta $p_\pm^\mu$ reads
\begin{eqnarray}
\label{SKG}
{\mathcal M}_{p_+p_-}^{(s_\pm=0)} = -ie\int d^4x\, \left(\Phi_{p_-}^*\partial_t\Phi_{p_+}-\Phi_{p_+}\partial_t\Phi_{p_-}^*\right) V(r)
\end{eqnarray}
where the nuclear field is treated in first order as before. The total pair creation rate is obtained like in Eq.\,\eqref{R}, with the spin sum omitted.

Total pair creation rates in collisions of an ultrarelativistic proton with a superintense optical laser beam are shown in Fig.\,1, based on Eqs.\,\eqref{S} and \eqref{SKG}. A very steep dependence on the proton's Lorentz factor $\gamma$ is found. The Doppler-enhanced laser field strength in the proton frame $E\approx 2\gamma E_0$ reaches the Schwinger value (i.e. $\chi\approx 1$) for Lorentz factors around $\gamma\approx 3500$, corresponding to the presently available proton energies of 3.5 TeV at CERN-LHC \cite{LHC}. Near-critical field strengths are required for sizeable pair yields since the production rate scales like $R\propto\exp(-2\sqrt{3}E_S/E)$ in the parameter regime where $\chi\ll 1$ \cite{Milstein}. Due to the exponential asymptotic behaviour of the rate, the pair creation is often said to occur in the tunneling regime here. In the overcritical regime ($\chi\gg 1$) the rate dependence on the field becomes logarithmic $R\propto\ln[E/(2\sqrt{3}E_S)]$ \cite{Milstein}. In the intermediate interaction regime considered here ($\chi\sim 1$) no exact analytical rate expressions are known and thus numerical computations of the rate are needed.

With regard to spin effects, Fig.\,1 demonstrates that the proper prediction based on the Dirac theory leads to production rates which are significantly larger than the Klein-Gordon results. For example, at $\gamma=1000$ ($\gamma=3000$) the Dirac theory predicts a total lab-frame rate of $2.1\times 10^5$\,s$^{-1}$ ($2.8\times 10^9$\,s$^{-1}$), whereas the Klein-Gordon treatment yields $1.9\times 10^4$\,s$^{-1}$ ($4.0\times 10^8$\,s$^{-1}$). Note that the enhancement is larger than the statistical factor 4 suggested by the number of possible spin configurations in the final state.

\begin{figure}[h]
\begin{center}
\resizebox{7cm}{!}{\includegraphics{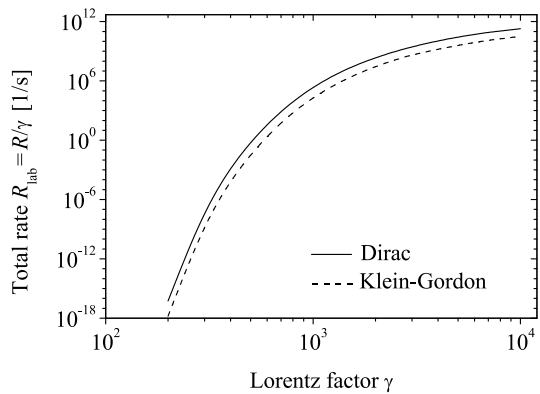}}
\caption{Total rates (with respect to the laboratory frame) for $e^+e^-$ pair creation in ultrarelativistic proton-laser collisions, as a function of the proton $\gamma$ factor. The parameters of the circularly polarized laser beam are $\omega_0=2.5$\,eV and $\xi=30$, corresponding to a field intensity of $10^{22}$\,W/cm$^2$. The solid and dashed lines show the predictions from the Dirac and Klein-Gordon theories, respectively.}
\end{center} 
\end{figure}

The total production rate in Eq.\,\eqref{R} may be decomposed into the contributions from different photon orders, $R=\sum_n R_n$. Formally this is achieved by expanding the periodic parts of the Volkov states into a Fourier series, resulting in an expression of the form \cite{MVG,Kaminski}
\begin{eqnarray}
|{\mathcal M}_{p_+p_-}|^2 = \sum_n T_n\,\delta(q_+^0 + q_-^0 -n\omega)
\label{n}
\end{eqnarray}
where the $T_n$ denote complex functions of the particle and laser field parameters. The argument of the $\delta$ function reflects the law of energy conservation in the process. Note that the leptons are created inside the strong field with their laser-dressed quasi-energies $q_\pm^0$ \cite{LL}. The corresponding laser-dressed mass amounts to  $m_\ast = m\sqrt{1+\xi^2}$. The typical number of laser photons absorbed is of the order of $n\sim 2m_\ast \xi/\omega$, with $\omega\approx 2\gamma\omega_0$, and varies over a broad range \cite{MVG}. This leads to a quasi-continuous number distribution, as Figure\,2 shows.

\begin{figure}[h]
\begin{center}
\resizebox{7.5cm}{!}{\includegraphics{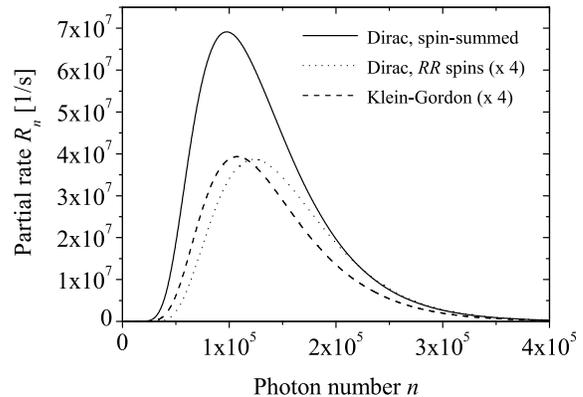}}
\caption{Distribution of the number of laser photons contributing to the pair creation rate, according to the Dirac and Klein-Gordon theories, as indicated. The laser parameters are as in Fig.\,1 and the proton Lorentz factor amounts to $\gamma=3000$.
The photon number distribution is equivalent to the spectrum of the total pair energy in the proton frame [see Eq.\,\eqref{n}]. In addition to the total rates (solid and dashed lines, respectively), the dotted line shows the contribution to the Dirac rate from pairs of right-handed particles; see Sec.\,III. The dashed and dotted curves are enhanced by a factor of 4.}
\end{center}
\end{figure}

\section{Spin-resolved production rates}
From the ordinary Bethe-Heitler effect by a single photon it is known that characteristic correlations between the electron and positron spins $s_\pm$ exist. For example, the faster one of the created leptons is much more likely to be polarized in the same sense as the circularly polarized photon \cite{McVoy,Olsen,beam}.

In this section, we analyze spin correlations in the nonlinear Bethe-Heitler effect in strong laser fields by determining the contributions from the various spin combinations to the total rate. To this end, we apply the spin projection operator of the Dirac theory
\begin{eqnarray}
\Sigma(s)=\frac{1}{2}\left(\mathds1+\gamma^5 \slashed{s}\right)\,,
\label{eq:spinproject}
\end{eqnarray}
with the spin four-vector $s^\mu=(s^0,\mbox{\boldmath$s$})$ satisfying the relations $s^2=-1$ and $(sp)=0$. We will consider the particular spin basis describing right-handed ($R$) and left-handed ($L$) particles, i.e.
\begin{eqnarray}
s_{R,L}^0 = \pm\beta_p \gamma_p\ , \qquad \boldsymbol s_{R,L} = \pm\gamma_p
\frac{\boldsymbol p}{|\boldsymbol p|}\ ,
\label{eq:spinRL}
\end{eqnarray}
where $\gamma_p=p^0/m$ and $\beta_p=(1-\gamma_p^{-2})^{1/2}$.

We note that, as long as the leptons move inside the field, their spin is precessing (see also \cite{Walser}). Formally this arises from the matrix-valued factor $\sim \slashed{k}\slashed{A}$ in Eq.\,\eqref{Volkov} which, upon multiplication, leads to a periodic rotation of the free Dirac spinor $u^{(\pm)}_{p,s}$. After the particles have left the laser field, the spin quantum numbers in Eq.\,\eqref{eq:spinRL} can be measured. The corresponding spinor wavefunctions then become eigenstates of the helicity operator $\hat{\boldsymbol S}\cdot \boldsymbol p/|\boldsymbol p|$ with the eigenvalues $\pm1/2$. Here, $\hat{\boldsymbol S}$ denotes the Dirac spin operator \cite{LL}. We recall that the helicity of a massive particle is not Lorentz invariant but depends on the frame of reference. Our results shown below refer to the rest frame of the proton.

With the help of the spin projection operator, the squared creation amplitude for a particular spin configuration may be expressed as a trace,
\begin{align}
\Bigl|{\mathcal M}_{p_+p_-}^{s_+s_-}\Bigr|^2 \propto
\text{Tr} \biggl(
&\Gamma
\left[
\left(\frac{\slashed{p}{}_+-m}{2m}\right)
\left(\frac{1+\gamma^5\slashed{s}{}_+}{2}\right)\right]
\notag\\
& \overline{\Gamma}
\left[
\left(\frac{\slashed{p}{}_- + m}{2m}\right)
\left(\frac{1+\gamma^5\slashed{s}{}_-}{2}\right)
\right]\biggr)\label{eq:trace}
\end{align}
with
\begin{eqnarray}
\Gamma = \left(1-\frac{e\slashed{A}\slashed{k}}{2(kp_+)}\right) \gamma^0
\left(1+\frac{e\slashed{k}\slashed{A}}{2(kp_-)}\right) \,.
\end{eqnarray}

Figure\,3 shows the relative contributions $R^{s_+s_-}/R$ of the various spin constellations to the total pair creation rate. Since the pair is created by the absorption of a very large number of circularly polarized photons of the same helicity, it is tempting to expect that both leptons are produced preferably with the photons' helicity. The analysis of the ordinary Bethe-Heitler process by single-photon absorption, however, led to the conclusion, that this intuitive expectation based on angular momentum conservation fails because of subtle quantum interference effects \cite{McVoy,LLpol}.

In the present situation, we find that the dominant contribution always stems from pairs in opposite spin states. Both combinations $e^+_Re^-_L$ and $e^+_Le^-_R$ are equally probable, since the square of the underlying matrix element is symmetric under particle exchange. This property is also known from the ordinary Bethe-Heitler process \cite{Olsen}. In comparison, the production of pairs of equal-handed leptons is significantly suppressed in this interaction regime. The probability for a right-handed pair is slightly larger than for a left-handed pair, which may be considered a consequence of the field helicity. While being smallest, the contribution from left-handed pairs nevertheless is much more sizeable than in pair creation by single-photon absorption, where it is practically negligible. Note, moreover, that the contributions from equal-handed leptons grow with increasing proton energy.

We point out that the contributions $R^{s_Rs_R}$ and $R^{s_Ls_L}$, respectively, from equal-handed pairs practically coincide with the total rate predicted by the spinless Klein-Gordon theory. Indeed, the ratio between the total Klein-Gordon and Dirac rates, $\rho=R_{\rm KG}/R_{\rm Dirac}$, amounts to $\rho\approx 0.09$ at $\gamma=1000$ and $\rho\approx 0.14$ at $\gamma=3000$ (see the corresponding numbers given in Sec.\,II). This observation is further corroborated by an additional calculation at $\gamma=300$ where $R^{s_Rs_R}/R\approx R^{s_Ls_L}/R\approx 0.04\approx \rho$ is obtained.
The Klein-Gordon prediction and the results for equal-handed spins according to the Dirac theory moreover give rise to very similar photon number distributions as well (see Fig.\,2).

\begin{figure}[h]
\begin{center}
\resizebox{8cm}{!}{\includegraphics{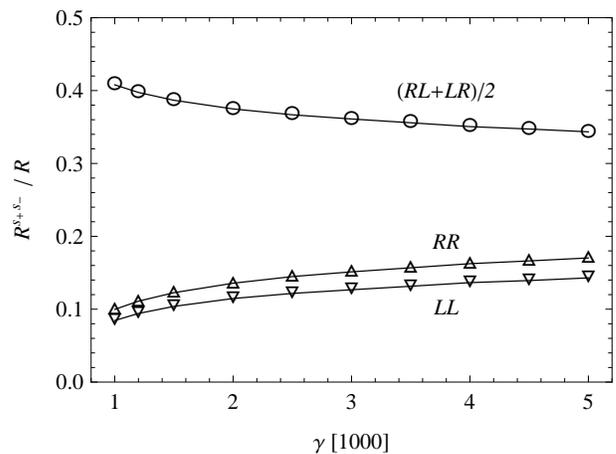}}
\caption{Relative contributions to the total pair creation rate from the different electron-positron spin configurations [see Eq.\,\eqref{eq:spinRL}], with respect to the proton frame. The laser parameters are as in Fig.\,1 and the proton energy varies from $\approx$\,1--5\,TeV. Solid lines connect our numerical data which are indicated by the open circles and triangles.}
\end{center}
\end{figure}

\section{Transfer of helicity and degree of polarization}

The ordinary Bethe-Heitler process by a single photon of circular polarization and high energy allows for a rather efficient helicity transfer to one of the produced leptons. The degree of polarization of each particle depends on its energy and, in the limit $\omega\gg m$, is given by \cite{Olsen}
\begin{equation}
P=\frac{dR_R-dR_L}{dR_R+dR_L}=\frac{4x-1}{4x^2-4x+3}\ .
\label{eq:pol}
\end{equation}
Here, $R_R\equiv R^{s_Rs_R}+R^{s_Rs_L}$ and $R_L\equiv R^{s_Ls_L}+R^{s_Ls_R}$, and $dR_{R,L}$ denote the corresponding singly differential rates with respect to the energy of the particle. The latter is expressed as $p^0=x\omega$, with $0\le x\le 1$, and refers to the nuclear frame of reference. 

For $p^0\approx\omega$, Eq.\,\eqref{eq:pol} implies $P\approx 1$. I.e., when almost the complete photon energy is transfered to one of the leptons, then this particle also carries the photon helicity (whereas the other particle of the pair is weakly polarized in the opposite sense, $P\approx-1/3$). Note that the high-energy lepton fulfills a dispersion relation $p^0\approx |{\boldsymbol p}|$ similar to a massless particle and is emitted under a narrow angle into forward direction. It thus closely resembles the absorbed photon in terms of the kinematic and polarization properties.

It is interesting to compare the degree of polarization \eqref{eq:pol} with the corresponding one resulting from a highly nonlinear Bethe-Heitler process involving the absorption of thousands of laser photons. Let us consider pair creation in proton-laser collisions at the parameters $\gamma=3000$, $\omega_0=2.5$\,eV and $\xi=30$. Then the photon energy in the proton frame amounts to $\omega=15$\,keV. The solid line in Fig.\,4 shows the degree of polarization for the dominant photon order $n=10^5$. The total amount of energy absorbed in this process is $n\omega=1.5$\,GeV. The degree of polarization from this partial rate is compared to the one obtained from the single-photon Bethe-Heitler process at the photon energy $\omega=1.5$\,GeV (dashed line), according to Eq.\,\eqref{eq:pol}. Both degrees of polarization are qualitatively distinct. The single photon process leads to a dominant production of right-handed particles over a broad energy range. Contrary to that, in the nonlinear case the degree of polarization is very close to zero for a wide range of medium energies. A significant degree of polarization is only realized at the edges of the energy spectrum. Interestingly, when one of the leptons absorbs almost the complete photon energy and becomes highly polarized, the low-energy lepton will obtain an almost equal degree of polarization in the opposite direction. Therefore, multiphoton absorption can in principle lead to a pair of particles which both are polarized.

However, the number of photons absorbed in the pair creation process is not fixed but varies over a broad range. The observable degree of polarization therefore is an average over all photon orders. Since the high-energy edges around $q^0\approx n\omega$ overlap with the energy spectra of neighboring photon orders, the degree of polarization will be less pronounced for the total process. We may thus conclude that the absorption of a large number of photons, although all of them carry the same helicity, seems to blur the effects that lead to high degrees of polarization in the single-photon process (see also \cite{Tsai}). 

\begin{figure}[h]
\begin{center}
\resizebox{8cm}{!}{\includegraphics{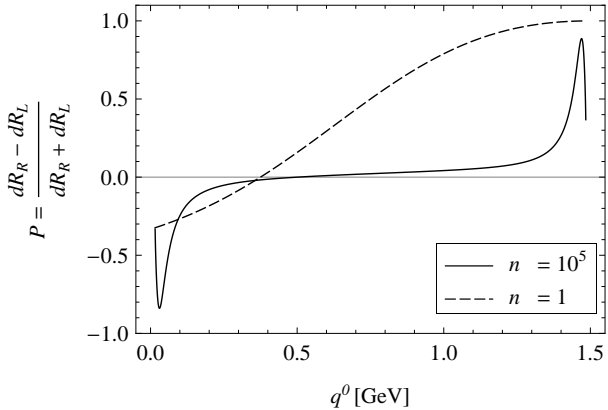}}
\caption{Degree of longitudinal polarization of one of the created particles for a single photon process (dashed line) in comparison with a highly nonlinear process involving $n=10^{5}$ laser photons (solid line). In the latter case, the collision parameters are the same as in Fig.\,2. An identical amount of energy (and momentum) is absorbed in both processes. The degree of polarization is plotted against the particle's quasi-energy in the proton frame. In the single-photon case (where $\xi\ll 1$ is assumed) the latter coincides with the free energy $p^0$.} 
\end{center} 
\label{fig:tunneling_pol}
\end{figure}

\section{Conclusion}

Spin effects in the nonlinear process of $e^+e^-$ pair creation in collisions of a highly relativistic proton beam with a super-intense optical laser pulse of circular polarization were considered. It was shown, that the leptonic spin degree of freedom substantially enhances the total pair creation rate, as compared to the prediction from the spinless Klein-Gordon theory. The spin states of the pairs are strongly correlated, favoring the creation of particles with opposite helicities in analogy with the linear Bethe-Heitler process. Nevertheless, only low degrees of polarization arise because of the very broad number distribution of laser photons absorbed. The contributions from pairs of equal helicity exhibit an interesting correspondance with the Klein-Gordon results.

Our predictions could be tested experimentally at CERN-LHC when a petwatt-class table-top laser system is operated in conjunction with the accelerator beamline. While the main goal of the LHC is the search for physics beyond the standard model, corresponding studies of strong-field QED phenomena might be conducted in a casual manner along the way, or at a later stage after the LHC has accomplished its major tasks.

\section{Acknowledgements}
Useful input by K. Z. Hatsagortsyan is gratefully acknowledged.


\begin{thebibliography}{99}
\bibitem{LLpol} An elegant general consideration of polarization effects (with respect to bremsstrahlung) may be found in chapter 93 of Ref.\,\cite{LL}.
\end{thebibliography}
\end{document}